\documentclass{article}       
\usepackage{emod}                
\usepackage{graphicx}  
\usepackage{color}     
\begin{document}
%
%
\branch{C}   
%
\title{The Road Ahead}
\author{S.~Willenbrock\inst{1,2}}
\institute{Department of Physics, University of Illinois at Urbana-Champaign,
1110 W.~Green Street, Urbana, IL 61801 \and Kavli Institute for Theoretical
Physics, University of California, Santa Barbara, CA 93106 \\e-mail:
willen@uiuc.edu}
\maketitle
\begin{abstract}
I describe the surrounding landscape on the road to the CERN Large Hadron
Collider. I revisit the milestones of hadron-collider physics, and from them
draw lessons for the future.  I recall the primary motivation for the journey
--- understanding the mechanism of electroweak symmetry breaking --- and
speculate that even greater discoveries may await us.  I review the physics
that we know beyond the standard model --- dark matter, dark energy, and
neutrino masses ---  and discuss the status of grand-unified theories. I list
the reasons why the Higgs boson is central to the standard model as well as to
physics beyond the standard model.
\end{abstract}
It's been a rough road for the hadron-collider community over the past
decade.  We've witnessed the death of the Superconducting Super Collider (SSC),
the big brother of the CERN Large Hadron Collider (LHC), in 1993.  We've
experienced delays in the schedules of both the Fermilab Tevatron and the
LHC.  The luminosity of Run II of the Tevatron has turned on more slowly than
desired.  I'm sure each of you can add your own personal frustrations to this
list. In addition, some of the biggest discoveries have occurred in other
areas of particle physics, and while we applaud these advances, it's hard not
to feel a twinge of jealousy.

It's important to remember that we've also had many successes, and there are
good reasons for optimism. Let's begin by looking backwards.

\section{The view in the rear-view mirror}
The CERN Super proton antiproton Synchrotron (S$p\bar p$S) was designed in the
mid 1970's to discover the $W$ and $Z$ bosons.  It succeeded in doing so in
1983 \cite{Arnison:rp,Arnison:1983mk,Banner:1983jy,Bagnaia:1983zx}, earning
the 1984 Nobel Prize in physics for C.~Rubbia \cite{Rubbia:pv} and S.~van der
Meer \cite{VanDerMeer:qh}.  The Fermilab Tevatron was designed to mass-produce
$W$ and $Z$ bosons, and that it has done spectacularly well.  The $W$-boson
mass has been measured at the Tevatron to be $M_W = 80.454 \pm 0.059$ GeV, an
accuracy of less than $0.1\%$.  We anticipate an accuracy of 30 MeV or less in
Run II of the Tevatron.  In the case of both the S$p\bar p$S and the Tevatron,
we delivered what we promised.

\begin{figure}
\begin{center}
\includegraphics[width=3in]{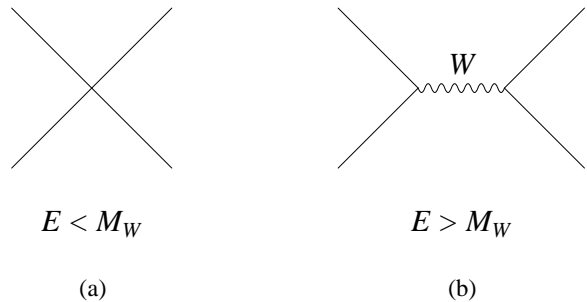}
\end{center}
\caption{(a) The effective theory of the weak interaction at energies less
than the $W$ mass is the Fermi theory; (b) at energies above the $W$ mass, the
effective theory is a spontaneously-broken $SU(2)_L\times U(1)_Y$ gauge
theory.}\label{fermitheory}
\end{figure}

From a theoretical perspective, we discovered the ``weak scale,'' $M_W$.  At
energies less than $M_W$, the effective theory of the weak interaction is the
Fermi theory, in which four fermions interact at a point, as shown in
Fig.~\ref{fermitheory}(a). At energies above $M_W$, the effective theory of the
weak interaction is a spontaneously-broken $SU(2)_L\times U(1)_Y$ gauge
theory, in which fermions interact by exchanging $W$ and $Z$ bosons, shown in
Fig.~\ref{fermitheory}(b).  However, the mechanism of spontaneous symmetry
breaking is unspecified.  It is appropriate to refer to $M_W$ as the ``weak
scale,'' because that is the energy at which one moves from one effective
theory to another.

We also delivered {\em more} than we promised.  Probably the most dramatic
discovery at the Tevatron (thus far) is the top quark.  We suspected that the
top quark exists ever since the discovery of the $b$ quark in 1977
\cite{Herb:ek}, but we did not know its mass.  By 1980 we had a lower bound of
$m_t>15$ GeV, which climbed to 23 GeV by 1984 \cite{Behrend:1984ub}. At any
given time it was commonly believed that the top quark lies just a few GeV
above the present lower bound.  The UA1 collaboration at the S$p\bar p$S
reported a signal for a top quark of mass around 40 GeV in 1984, but it proved
to be a red herring \cite{Arnison:1984iw}.  By 1987 we had an upper bound of
200 GeV on the top-quark mass from precision electroweak data
\cite{Amaldi:1987fu}.

\begin{figure}[t]
\begin{center}
\includegraphics[width=\textwidth]{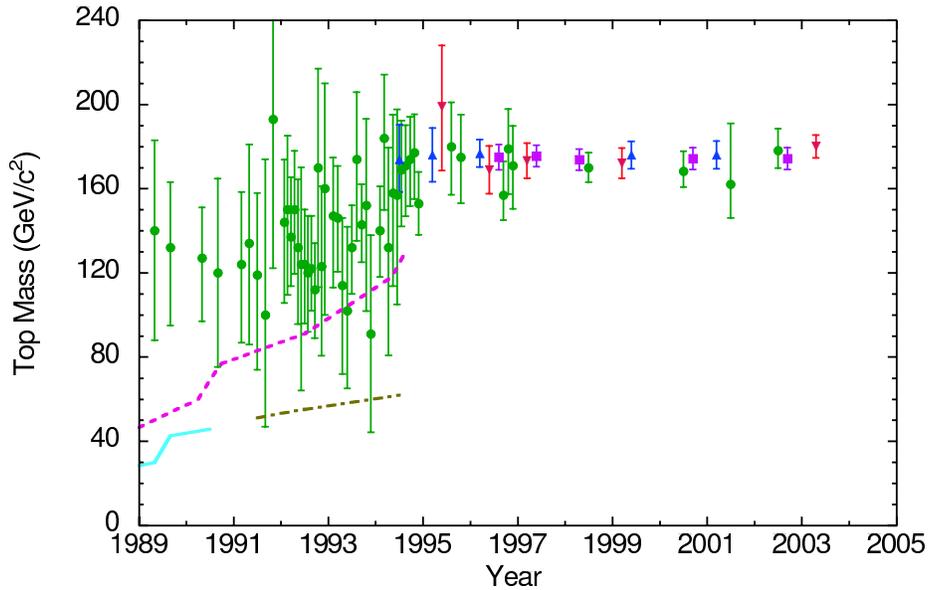}
\end{center}
\caption{($\diamondsuit$) Indirect bounds on the top-quark mass from precision
electroweak data. ($\sqcap\!\!\!\!\sqcup$) World-average direct measurement of
the top-quark mass; ($\bigtriangleup$) CDF and ($\bigtriangledown$) D0
measurements.  Lower bounds from $p\bar p$ (dashed) and $e^+e^-$ (solid)
colliders.  Updated by C.~Quigg from Ref.~\cite{Quigg:fy}.} \label{Top2003}
\end{figure}

The precision electroweak data became much more precise in 1989 with the
advent of the CERN Large Electron-Positron (LEP) collider and the Stanford
Linear Collider (SLC).  I show in Fig.~\ref{Top2003} the top-quark mass
extracted from precision electroweak analyses from 1989 to the present.  As
the data became more precise, and the lower bound from the Tevatron continued
to increase, it became clear that the top quark is much heavier than expected.
This culminated in the discovery of the top quark in 1995
\cite{Abe:1995hr,Abachi:1995iq} with a mass around 175 GeV, in the range
anticipated by precision electroweak analyses.

The top quark was not a central motivation for the S$p\bar p$S nor the
Tevatron, but it proved to be one of the most important discoveries at these
machines. We anticipated the existence of the top quark, and we used precision
electroweak data to successfully determine the allowed mass range. This was a
great success, and was part of the reason the 1999 Nobel Prize was awarded to
G.~'t Hooft \cite{'tHooft:xn} and M.~Veltman \cite{Veltman:xp}. However, the
large mass of the top quark was a real surprise.

Along the way, we developed new experimental techniques that could not have
been dreamt of when the S$p\bar p$S and the Tevatron were being designed. For
the top quark, the ability to tag $b$ quarks using a silicon vertex detector
(SVX) is a very powerful tool.  It wasn't until 1983 that we discovered that
the $b$ quark has a surprisingly long lifetime
\cite{Fernandez:1983az,Lockyer:1983ev} that might allow one to detect a
secondary vertex from $b$ decay.  By the mid 1980's it was well appreciated
that this could be used to tag $b$ jets at hadron colliders, but the
anticipated efficiency was quite low \cite{Brau:1988ic}. When the top quark was
discovered in 1995, we learned that the efficiency for SVX $b$-tagging is as
much as $50\%$, yet another surprise in the saga of the top quark.

\section{The LHC}
\begin{figure}
\begin{center}
\includegraphics[width=3.5in]{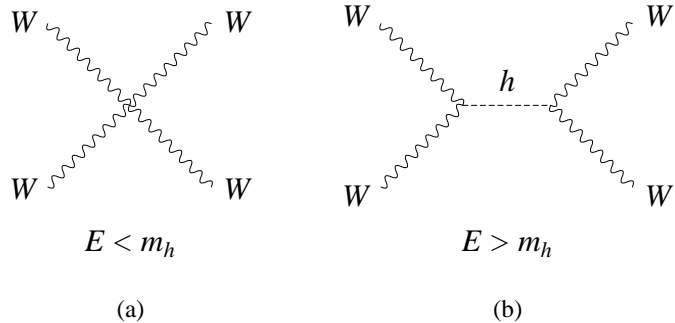}
\end{center}
\caption{(a) The effective theory of the weak interaction at energies less
than the Higgs mass is the spontaneously-broken $SU(2)_L\times U(1)_Y$ gauge
theory; (b) at energies above the Higgs mass, the effective theory is the
standard model.}\label{higgstheory}
\end{figure}

The central motivation for the LHC is to discover the mechanism of electroweak
symmetry breaking.  Recall that this mechanism is not specified in the
spontaneously-broken $SU(2)_L\times U(1)_Y$ gauge theory.  The amplitude for
the scattering of $W$ bosons in that theory is shown in
Fig.~\ref{higgstheory}(a). In the standard model, one introduces a Higgs field
that acquires a vacuum expectation value and breaks the electroweak symmetry.
This results in a new particle in the theory, the Higgs boson.  Thus there is
an additional diagram, involving the exchange of the Higgs boson, that
contributes to the amplitude for $W$-boson scattering, shown in
Fig.~\ref{higgstheory}(b). For energies less than the Higgs-boson mass, the
effective theory is the spontaneously-broken $SU(2)_L\times U(1)_Y$ gauge
theory.  The effective theory above the Higgs-boson mass includes the Higgs
field, which provides the mechanism for electroweak symmetry breaking. Thus it
is appropriate to refer to the Higgs-boson mass as the ``scale of electroweak
symmetry breaking'' in the standard model.

In the spontaneously-broken $SU(2)_L \times U(1)_Y$ gauge theory without a
specific mechanism for symmetry breaking, the gauge symmetry is realized
nonlinearly.  When a specific mechanism is introduced, the gauge symmetry is
realized linearly.  Thus the general definition of the ``scale of electroweak
symmetry breaking'' is the scale above which the gauge symmetry of the
effective theory is realized linearly.  The standard model is a specific
example of this general definition.

\begin{figure}[t]
\begin{center}
\includegraphics[width=.6\textwidth]{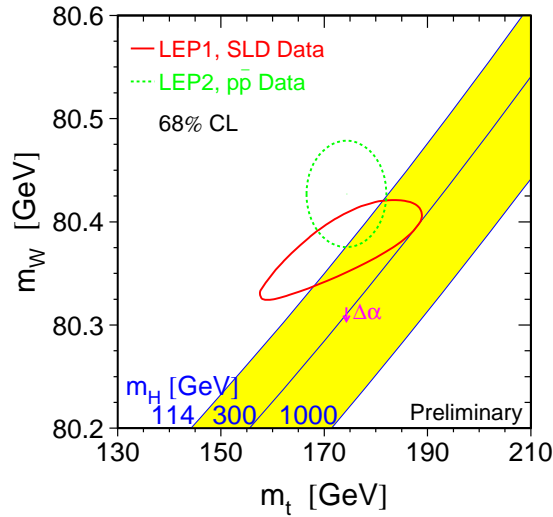}
\end{center}
\caption{Lines of constant Higgs mass on a plot of $M_W$ {\it vs.}~$m_t$.  The
dashed ellipse is the $68\%$ CL direct measurement of $M_W$ and $m_t$.  The
solid ellipse is the $68\%$ CL indirect measurement from precision electroweak
data.  From http://lepewwg.web.cern.ch/LEPEWWG.} \label{w03_mt_mw_contours}
\end{figure}

\begin{figure}
\begin{center}
\includegraphics[width=.75\textwidth]{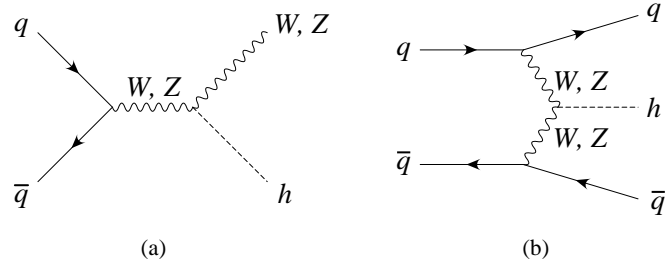}
\end{center}
\caption{Higgs-boson production (a) in association with a weak vector boson,
(b) via weak-vector-boson fusion.} \label{higgsproduction}
\end{figure}

\begin{figure}
\begin{center}
\includegraphics[width=.6\textwidth]{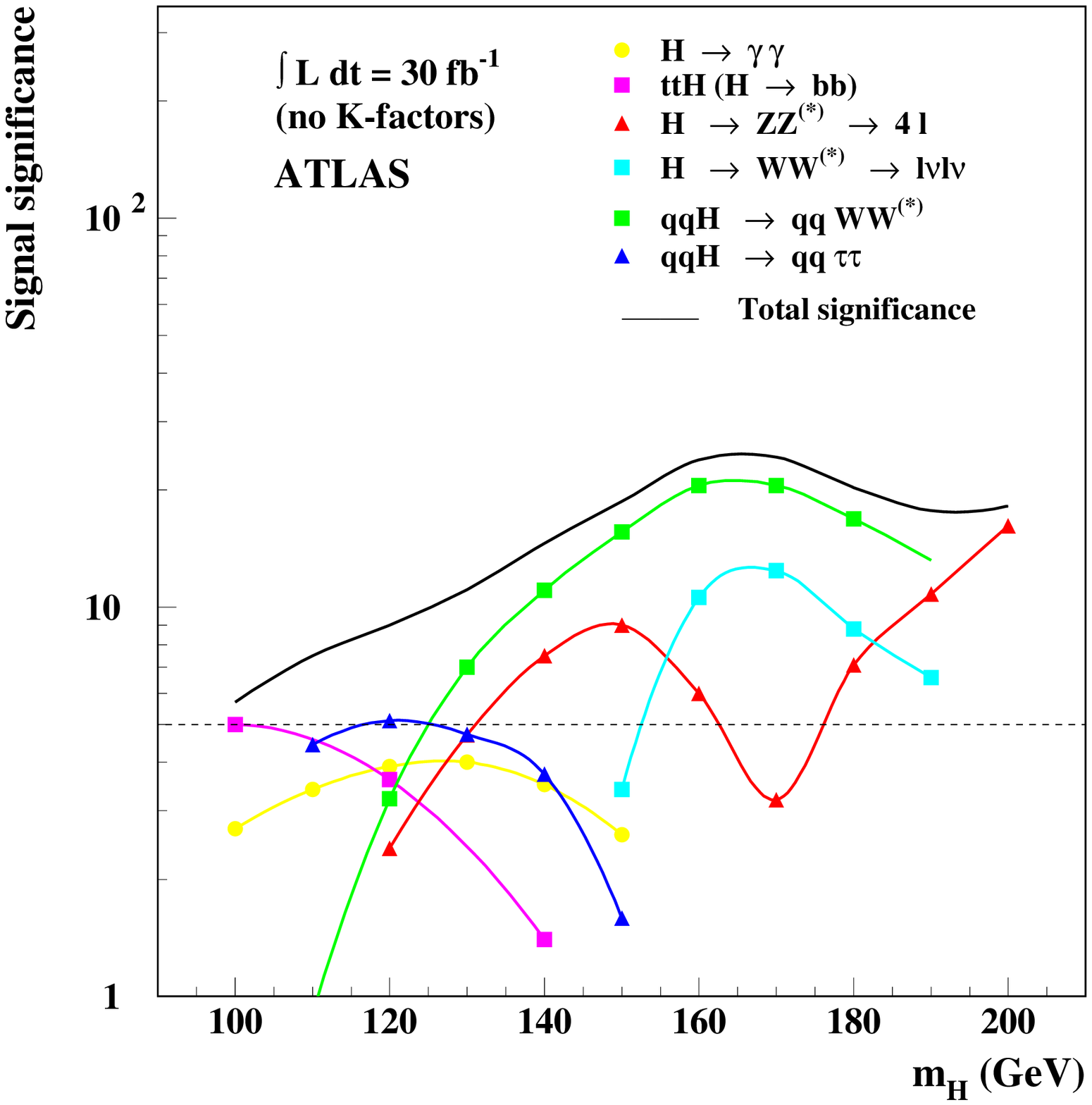}
\end{center}
\caption{$S/\sqrt B$ for a variety of Higgs-boson production and decay
channels at the LHC {\it vs.}~the Higgs mass.} \label{higgs_vbf_30}
\end{figure}

Let's consider the possibility that the standard Higgs model is correct.  Just
as precision electroweak data honed in on the allowed range of the top-quark
mass, we can use this data to determine the allowed range for the Higgs-boson
mass. The precision electroweak data are summarized on a plot of the $W$ mass
{\it vs.} the top-quark mass, shown in Fig.~\ref{w03_mt_mw_contours}.  Lines of
constant Higgs mass are drawn on this plot.  The elongated ellipse represents
the precision electroweak data, assuming the standard Higgs model. The dashed
ellipse indicates the direct measurement of the $W$ mass and the top-quark
mass.  The most striking thing about his plot is that the two ellipses overlap
near the lines of constant Higgs mass. This did not have to happen: these two
ellipses could have ended up anywhere on this plot (or even off of it), and
they did not have to overlap.  This indicates that the standard Higgs model is
at least a good approximation to reality.  Furthermore, the ellipses overlap
near the lines of constant Higgs mass corresponding to small values of the
Higgs mass. This indicates that the Higgs boson is not much heavier than the
present lower bound of $m_h>114.4$ GeV \cite{LEPHiggs}.

Such an intermediate-mass Higgs boson may be accessible in Run II at the
Tevatron via the associated production of the Higgs boson and a weak vector
boson, as shown in Fig.~\ref{higgsproduction}(a) \cite{Stange:ya}.  This is
remarkable because the Higgs boson was not at all a motivation for the
Tevatron.  This search channel only became feasible once we realized that we
could tag $b$ jets with high efficiency, since the Higgs boson decays
dominantly to $b\bar b$ in the Higgs-mass region of interest.  The discovery
of the Higgs boson via this process requires a lot of integrated luminosity
\cite{Amidei:1996dt,Carena:2000yx}. However, we should keep in mind that our
projections about the required luminosity for a given measurement are
sometimes too conservative.  A striking example is the projected accuracy in
the measurement of the top-quark mass made in the TeV-2000 study in 1996,
$\delta m_t=13$ GeV (with 70 pb$^{-1}$ of data) \cite{Amidei:1996dt}. Just two
years later CDF and D0 measured the mass with a combined accuracy of 5.1 GeV
(with 100 pb$^{-1}$ of data) \cite{Demortier:1999vv}.  We greatly
underestimated the sensitivity of the measurement, despite the fact that at
the time we probably thought we were being optimistic.  The lesson is that
``data make you smarter'' \cite{FERMINEWS}.

Studies also make you smarter.  In the context of an intermediate-mass Higgs
boson, this is exemplified by the production of the Higgs boson via
weak-vector-boson fusion, shown in Fig.~\ref{higgsproduction}(b).  This process
was originally thought to be of interest primarily for a heavy Higgs boson,
$m_h > 2M_W$ \cite{Cahn:ip}. However, in recent years it has been realized
that it is also very important for an intermediate-mass Higgs boson
\cite{Rainwater:1997dg}. I show in Fig.~\ref{higgs_vbf_30} the signal
significance for an intermediate-mass Higgs boson at the LHC via a variety of
production and decay channels.  The importance of the weak-vector-boson
channels is evident from this plot.

\section{The view through the sunroof}
\begin{figure}
\begin{center}
\includegraphics[width=4in]{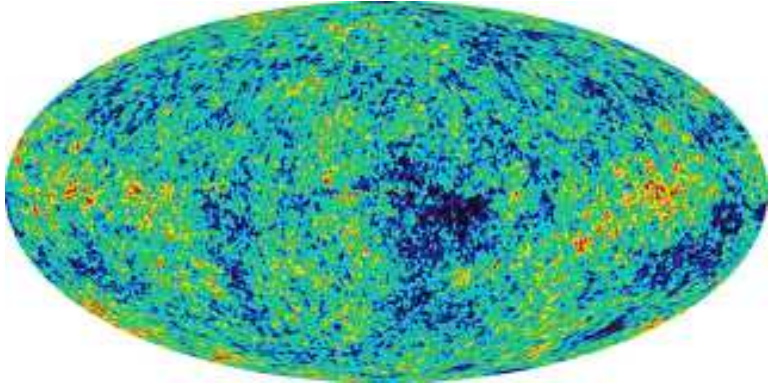}
\end{center}
\caption{Temperature fluctuations in the cosmic microwave background as seen
by WMAP.  From Ref.~\cite{Bennett:2003bz}.}\label{020598_ilc_640}
\end{figure}

Figure~\ref{020598_ilc_640} shows the WMAP data on the temperature fluctuations
in the cosmic microwave background radiation \cite{Bennett:2003bz}.
Cosmological parameters can be extracted from this data with impressive
precision.  The total energy density of the universe, in units of the critical
density, is very close to unity,
\begin{displaymath}
\Omega_{TOT}= \Omega_B+\Omega_{DM}+\Omega_{\Lambda} = 1.02\pm 0.02 \;,
\end{displaymath}
where $\Omega_B$, $\Omega_{DM}$, $\Omega_{\Lambda}$ are the baryon,
dark-matter, and dark-energy densities in units of the critical density,
\begin{eqnarray*}
\Omega_B&=&0.044\pm 0.004 \\
\Omega_{DM}&=&0.22\pm 0.04 \\
\Omega_{\Lambda}&=&0.73\pm 0.04 \;.
\end{eqnarray*}
We have known about the existence of dark matter for a long time, and now the
dark-matter density is known with good accuracy.  Dark energy has come and gone
throughout the decades, but now it looks like it is here to stay. In addition,
$\Omega_{TOT}\approx 1$ and other features of the data are consistent with
inflation.

Could we find dark matter at the LHC?  The discovery of dark matter was not one
of the original motivations for the LHC --- there is no mention of it in the
proceedings of the 1982 Snowmass study that gave birth to the SSC
\cite{Snowmass82} --- although it later became one of the goals of the project
\cite{Quigg:1985qm}.  However, the discovery of dark matter, which makes up
22\% of the universe, is potentially more exciting than understanding the
mechanism of electroweak symmetry breaking.

\begin{figure}
\begin{center}
\begin{picture}(420,420)
\put(-40,210){\includegraphics[width=7cm]{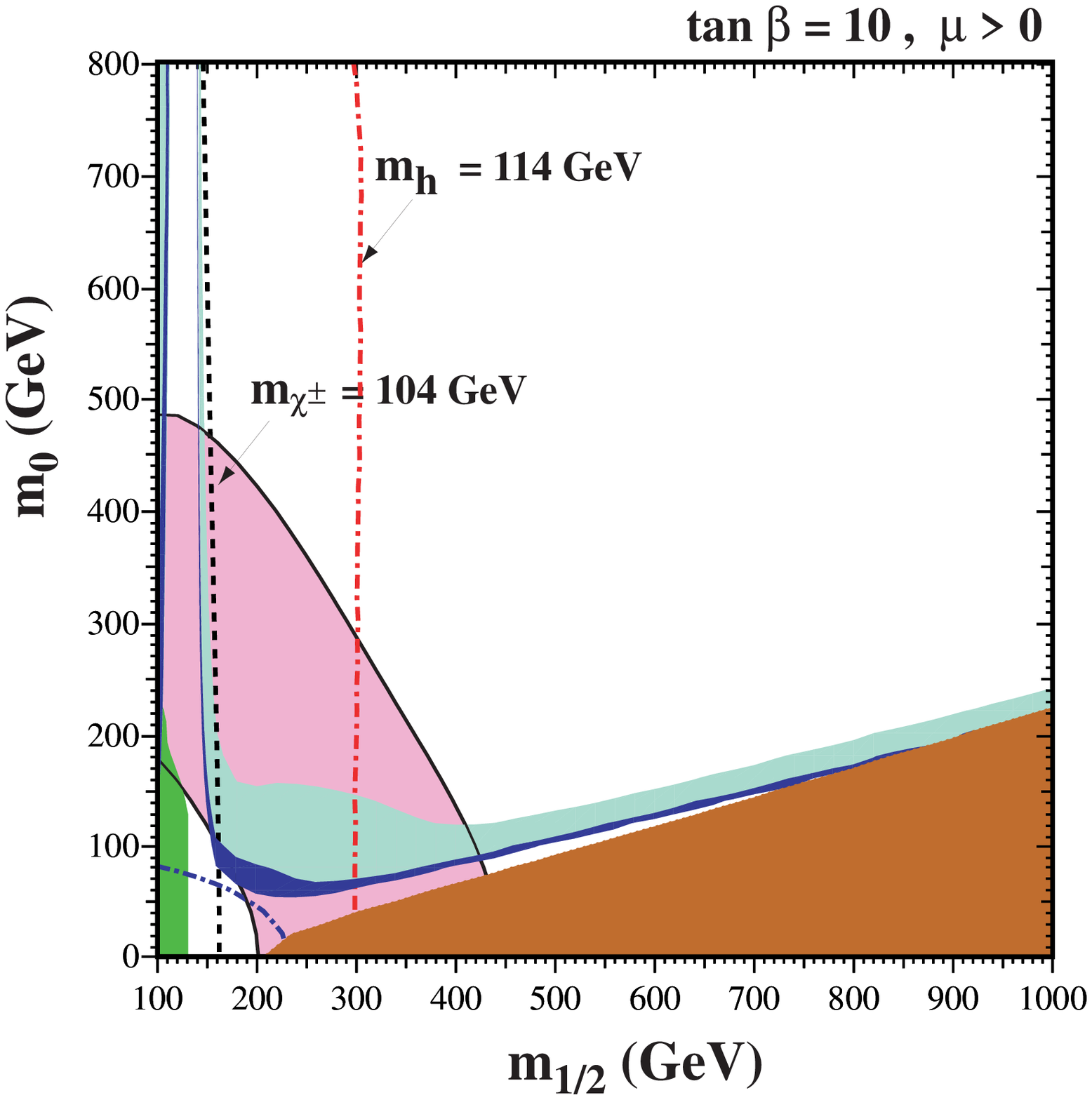}}
\put(170,210){\includegraphics[width=7cm]{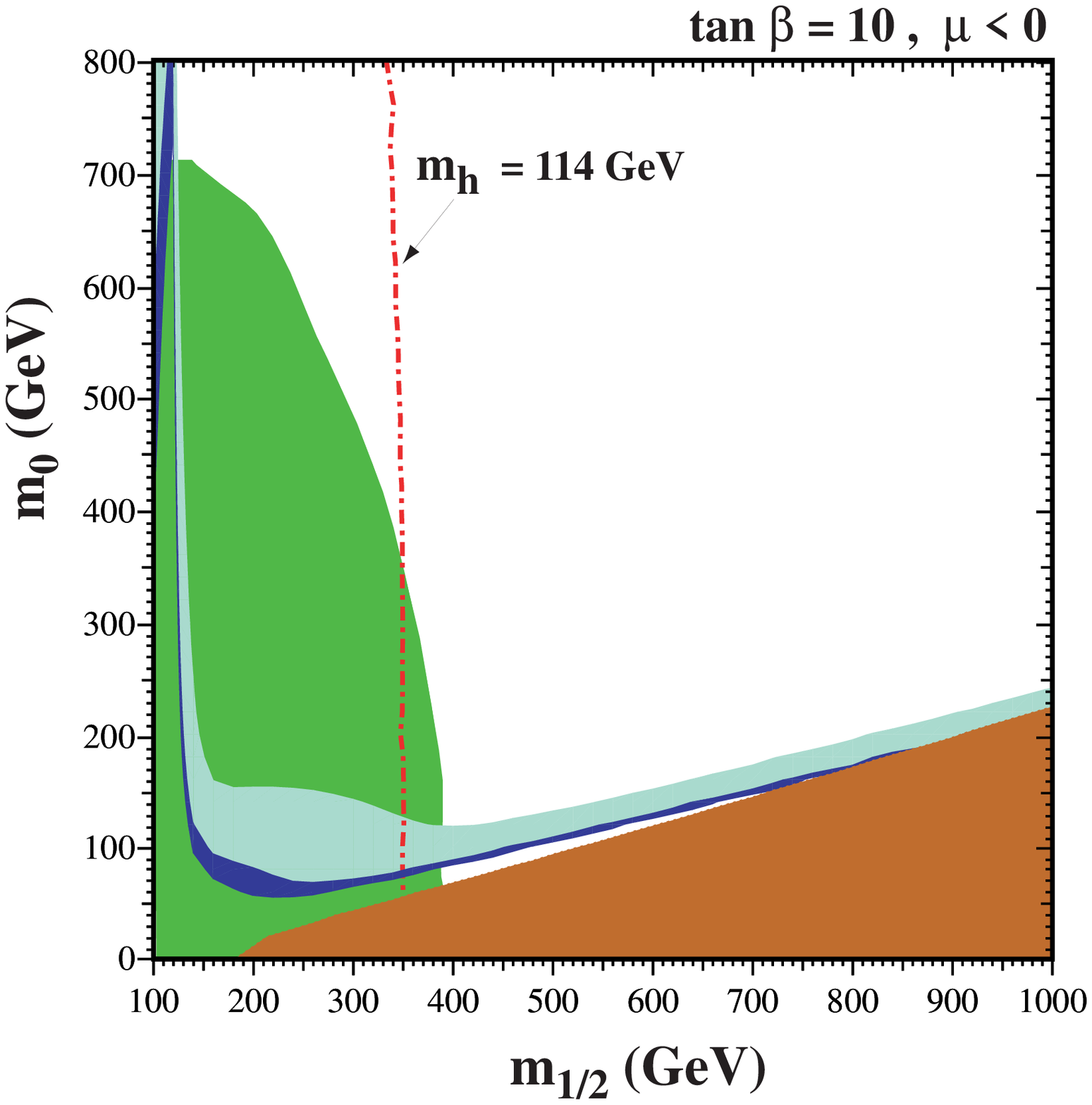}} \put(-40,
0){\includegraphics[width=7cm]{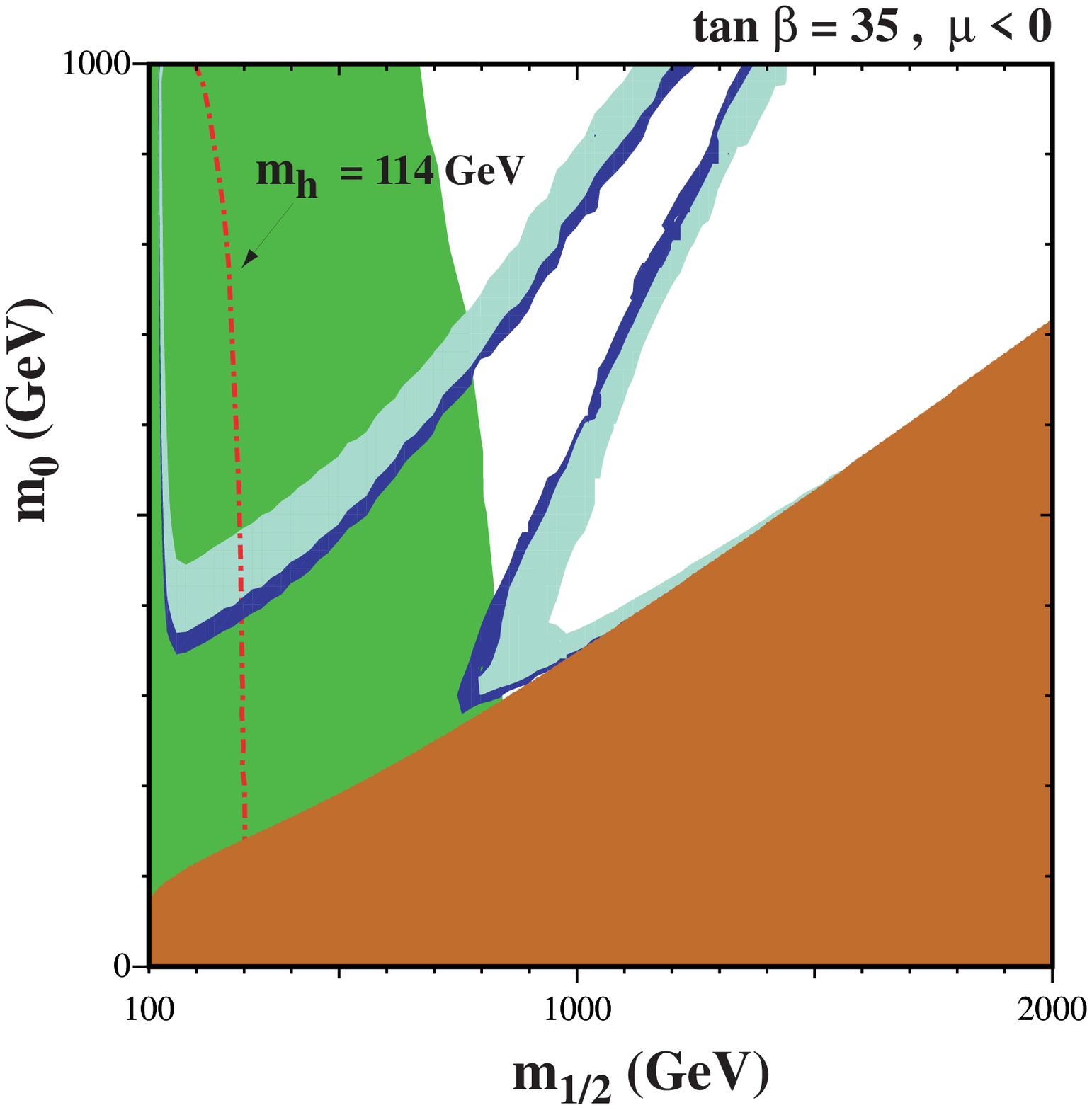}} \put(170,
0){\includegraphics[width=7cm]{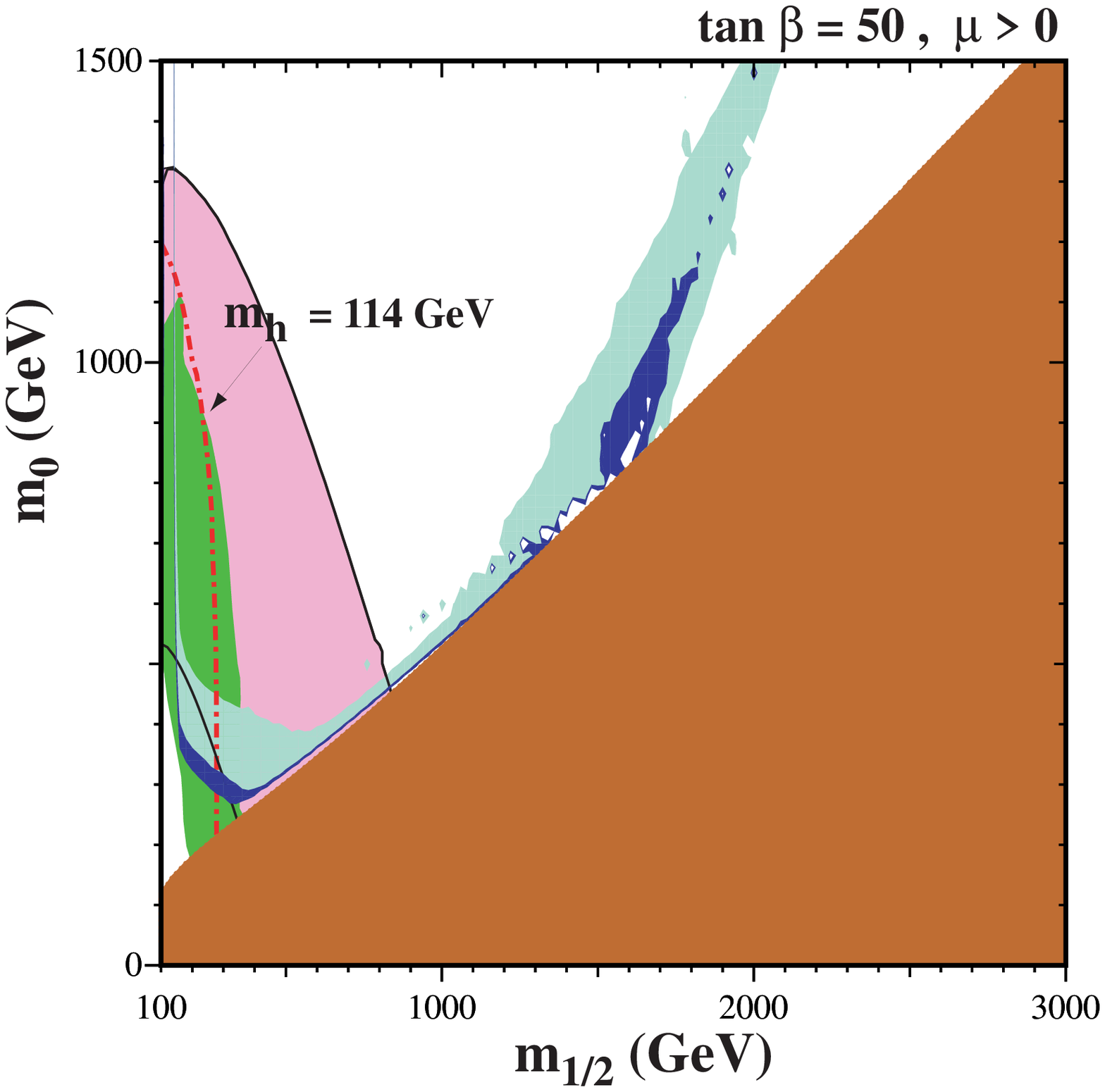}}
\end{picture}
\end{center}
\caption{The $(m_{1/2}, m_0)$ planes for (a) $\tan\beta = 10, \mu > 0$, (b)
$\tan\beta = 10, \mu < 0$, (c) $\tan\beta = 35, \mu < 0$, and (d) $\tan\beta =
50, \mu > 0$ ($A_0=0$ in all cases). In each panel, the region allowed by the
older cosmological constraint $0.1 \le \Omega_\chi h^2 \le 0.3$ has medium
shading, and the region allowed by the newer cosmological constraint $0.094 \le
\Omega_\chi h^2 \le 0.129$ has very dark shading. The disallowed region where
$m_{\tilde \tau_1} < m_\chi$ has dark (red) shading. The regions excluded by $b
\rightarrow s \gamma$ have medium (green) shading. (The regions of medium
(pink) shading in panels (a,d), formerly favored by $g_\mu-2$, are now
obsolete.) A dot-dashed line in panel (a) delineates the LEP constraint on the
$\tilde e$ mass and the contours $m_{\chi^\pm} = 104$~GeV ($m_h = 114$~GeV)
are shown as near-vertical black dashed (red dot-dashed) lines in panel (a)
(each panel). From Ref.~\cite{Ellis:2003cw}.}\label{susydark}
\end{figure}

Supersymmetry (SUSY) provides an attractive candidate for the dark matter, the
``neutralino,'' which is a linear combination of the photino, Zino, and
Higgsinos.  If it is the lightest supersymmetric particle, and $R$ parity is
conserved, then it is stable.  However, now that we know the dark-matter
density with good accuracy, it turns out that supersymmetry generically
produces {\em too much} dark matter.  Figure~\ref{susydark} shows the regions
of SUSY parameter space that are consistent with the dark-matter density,
before and after the WMAP data.  Before the WMAP data, there were large regions
allowed, but after the data there are only slivers of parameter space that
survive. This is noteworthy because these slivers represent regions in which
special coincidences occur that allow the dark matter to annihilate. In the
top two diagrams of Fig.~\ref{susydark}, the slivers extending towards the
right represent the case where the supersymmetric partner of the tau lepton is
nearly degenerate with the neutralino, so it is present in sufficient
abundance for co-annihilation to occur ({\it e.g.}, $\chi^0\tilde\tau_1\to
\gamma\tau$). In the bottom two diagrams, the slivers correspond to the
neutralino mass being close to half the mass of a Higgs boson, such that
annihilation occurs via the Higgs resonance.  In the first of these two
diagrams, there are two slivers
--- in the ``funnel'' between them, too much dark matter is annihilated.  This
is the most natural solution; one must be close to a Higgs resonance, but not
right on top of it, to get the correct relic abundance of dark matter.  Another
natural solution, at very large values of $m_0$ (not shown in the figures), is
the ``focus-point'' region \cite{Feng:2000gh}.

Dark energy is much harder to explain than dark matter.  Like dark matter, dark
energy was not anticipated.  As far as we can tell, dark energy is a constant
over space, and acts like a cosmological constant, $\Lambda$.  It is hard to
understand why $\Lambda \ll M_{P}^2$ ($M_P$ is the Planck scale), or even why
$\Lambda \ll v^2$ ($v$ is the Higgs-field vacuum expectation value). For many
years it was assumed that $\Lambda$ is exactly zero, and that we would some day
discover the mechanism that ensures that it vanishes.  Now we have the harder
problem of explaining why it is so small and yet not exactly zero
\cite{Carroll:2001xs}.

Another thing we learn by looking through the sunroof is that neutrinos, both
solar and atmospheric, oscillate.  This implies that neutrinos have a small
mass.  Like dark matter and dark energy, this is physics beyond the standard
model.

\begin{table}[b] \caption{The fermions of the first generation.  The
right-handed neutrino field, $N_R$, is not present in the standard model.}
\begin{center}\begin{tabular}[7]{cc}
$Q_L\equiv\left(\begin{array}{l}u_L\\d_L\end{array}\right)$&$\begin{array}{l}u_R\\d_R\end{array}$\\
\\
$L_L\equiv\left(\begin{array}{l}\nu_{L}\\e_L\end{array}\right)$&$\begin{array}{l}N_R\\e_R\end{array}$
\end{tabular}\end{center} \label{tab:sm}
\end{table}

\begin{figure}
\begin{center}
\includegraphics[width=4in]{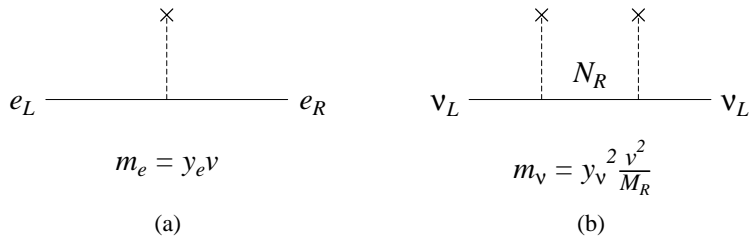}
\end{center}
\caption{(a) The electron acquires a Dirac mass via its coupling to the Higgs
field, $y_e$; (b) the neutrino acquires a Majorana mass via the square of its
coupling to the Higgs field, $y_\nu$, and an intermediate, heavy, right-handed
neutrino.}\label{neutrinomass}
\end{figure}

I show in Table~\ref{tab:sm} the fermions of the first generation.  I have
added a right-handed neutrino field, $N_R$, which is not present in the
standard model.  It is plausible that such a field exists --- why should all
the other left-handed fields have right-handed partners, but not the neutrino?
However, this field is special, because it is the only one that does not have
$SU(2)_L\times U(1)_Y$ interactions --- it is completely inert. Unlike the
other fields, which are forbidden from having a mass by the $SU(2)_L\times
U(1)_Y$ gauge symmetry, the field $N_R$ is allowed to have a mass, and
therefore we expect that it does.  The other fields acquire a mass, via their
coupling to the Higgs field, only when the $SU(2)_L\times U(1)_Y$ gauge
symmetry is spontaneously broken.  This is illustrated in
Fig.~\ref{neutrinomass}(a) for the electron, where the left- and right-handed
electron fields together acquire a Dirac mass proportional to their coupling
to the Higgs field times the Higgs vacuum-expectation value.

Since the right-handed neutrino field has a mass, it does not simply pair up
with the left-handed neutrino field to generate a Dirac mass.  Instead, it
generates a Majorana mass for the left-handed neutrino field, as shown in
Fig.~\ref{neutrinomass}(b).  This requires two interactions with the Higgs
field, so the neutrino mass is proportional to the square of the coupling to
the Higgs field times the square of the Higgs vacuum-expectation value,
divided by the mass of the right-handed neutrino field (which enters via its
propagator).  If $M_R$ is much greater than $v$, then the neutrino is very
light.  Taking $M_R$ to be around the scale of grand unification yields
neutrino masses in the range $10^{-5}-10^{2}$ eV \cite{Langacker:1980js},
consistent with what we know from neutrino oscillation experiments.  Thus we
anticipated neutrino masses from grand unification.  However, we did not
anticipate the large observed mixing angles, $\theta_{12}\approx 34^\circ$,
$\theta_{23}\approx 45^\circ$. Like the top quark, this is another example
where we anticipated the general framework, but not the details.

\section{Grand Unification}
\begin{figure}[p]
\begin{center}
\input{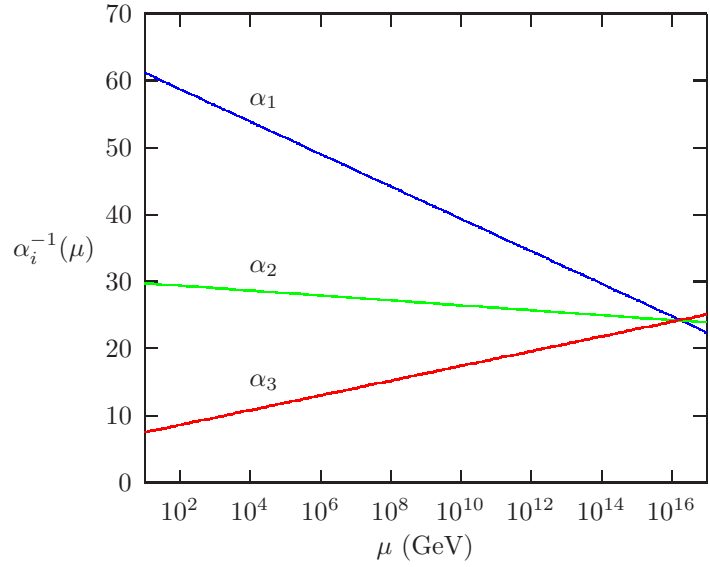}
\end{center}
\caption{Gauge coupling unification in the minimal supersymmetric standard
model.}\label{susy}
\end{figure}

\begin{figure}[p]
\begin{center}
\input{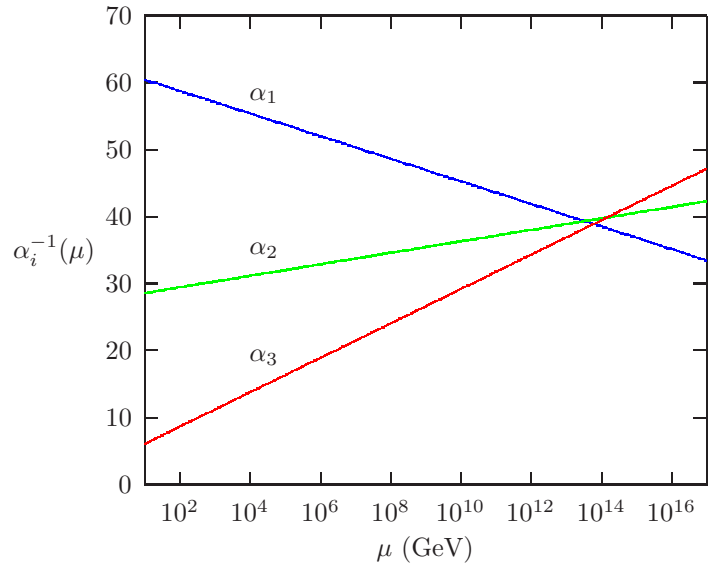}
\end{center}
\caption{Gauge coupling unification in the standard model with two Higgs
doublets plus their supersymmetric partners.  The same result is obtained in
the standard model with six Higgs doublets.  From
Ref.~\cite{Willenbrock:2003ca}.}\label{sm6}
\end{figure}

Since neutrino masses support the framework of grand unification, let's
consider the status of such theories.  The standard model fits neatly into
$SU(5)$ \cite{Georgi:sy}, but the couplings fail to unify at the grand-unified
scale.  As is well know, coupling unification is successful if one extends the
standard model to include supersymmetry in the minimal way
\cite{Ellis:1990wk,Amaldi:1991cn,Langacker:1991an}, as shown in
Fig.~\ref{susy}. What is less well known is that this is due entirely to the
extension of Higgs sector to include a second Higgs doublet and the
superpartners of the two Higgs doublets.  To illustrate this point, I show in
Fig.~\ref{sm6} the evolution of the gauge couplings obtained by adding just
the second Higgs doublet and the Higgs superpartners.  Exactly the same
evolution is obtained in the standard model with six Higgs doublets
\cite{Willenbrock:2003ca}. However, the unification scale is around $10^{14}$
GeV, which implies rapid proton decay in the $SU(5)$ model. One of the
attractive features of supersymmetry is that it not only allows for coupling
unification, it also pushes the unification scale up to around $10^{16}$ GeV,
making it safe from rapid proton decay \cite{Dimopoulos:1981yj}.

Once we accept the existence of the right-handed neutrino field, $N_R$, it is
attractive to consider $SO(10)$ as the grand-unified gauge group
\cite{Georgi:my,Fritzsch:nn}. The fermions of a single generation fit into the
$\bar 5+10+1$ representation of $SU(5)$, where the $1$ is the $N_R$ field,
which is simply tacked onto the theory.  In contrast, the fermions of a single
generation fill out the $16$ representation of $SO(10)$.  Thus $SO(10)$ is
more unified than $SU(5)$.  It is possible that $SO(10)$ is spontaneously
broken to $SU(5)$ at or above the grand-unified scale, in which case we are
led back to the $SU(5)$ scenario discussed above. However, there are other
possible symmetry-breaking patterns, which do not necessarily require
weak-scale supersymmetry.  A non-supersymmetric example is $SO(10)\to
SU(4)_c\times SU(2)_L\times SU(2)_R \to SM$ \cite{Pati:1974yy}. In order to
achieve coupling unification, there is an intermediate scale of symmetry
breaking between the weak scale and the grand-unified scale
\cite{Deshpande:1992au}. Since this intermediate scale is adjusted to yield
coupling unification, we lose the {\em prediction} of the weak mixing angle
that is one of the successes of the supersymmetric $SU(5)$ theory.

\begin{figure}[t]
\begin{center}
\includegraphics[width=4in]{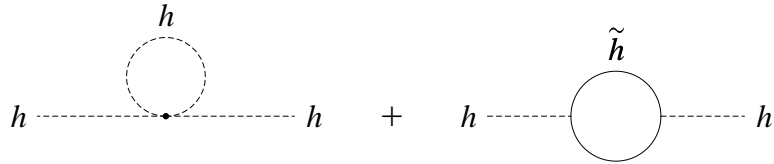}
\end{center}
\caption{Quadratic divergences from loops of particles and their
supersymmetric partners cancel in supersymmetric theories, exemplified here by
Higgs and Higgsino loops.}\label{susyloop}
\end{figure}

Another reason weak-scale supersymmetry is attractive is that it stabilizes
the hierarchy $m_h \ll M_U$, where $M_U$ is the unification scale (although it
does not explain why the hierarchy exists).  This stabilization results from
the cancellation of quadratically-divergent corrections to the Higgs-boson
mass from loops of particles and their supersymmetric partners, as shown in
Fig.~\ref{susyloop} \cite{Veltman:1980mj,Maiani:cx,Witten:nf}.  However,
weak-scale supersymmetry fails to stabilize the hierarchy $\Lambda \ll v^2$,
so it seems we are still missing a big part of the picture.

I believe that $SO(10) \to SU(5) \to SM$ with weak-scale supersymmetry is the
most attractive theory we've got, but it is unlikely that we have anticipated
all the details, just as we failed to anticipate the top-quark mass and the
neutrino mixing angles. The LHC will decide if weak-scale supersymmetry is
really an outpost on the path to grand unification, or simply a mirage.

\section{The Higgs Boson}

As I discussed above, the LHC was designed to discover the mechanism of
electroweak symmetry breaking.  The evidence suggests that this involves a
Higgs field (or fields) that acquires a vacuum-expectation value.  Here I
would like to list the intellectual reasons why the discovery of the Higgs
boson (or bosons) is so important:

\begin{itemize}

\item We have no experience in particle physics with a fundamental scalar
field, nor with a scalar field that acquires a vacuum-expectation value.  We
do have experience with a composite field that acquires a vacuum-expectation
value in QCD, $\langle \bar qq\rangle$, where $q$ is a quark field.  This
breaks the chiral symmetry of QCD down to isospin.  The analogue of this
mechanism for electroweak symmetry breaking is called Technicolor, and is an
alternative to a fundamental scalar field \cite{Weinberg:bn,Susskind:1978ms}.

\begin{figure}
\begin{center}
\includegraphics[width=2in]{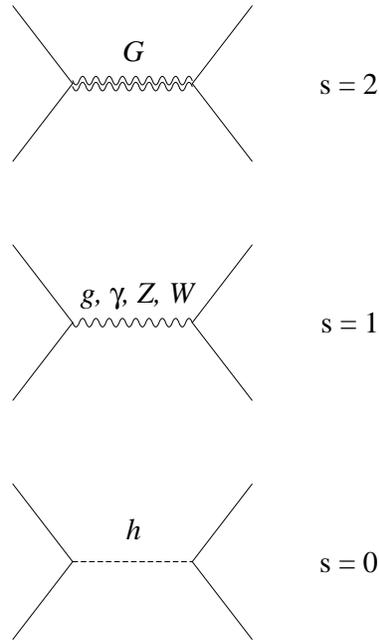}
\end{center}
\caption{Like the graviton and the gauge bosons of the standard model, the
Higgs boson mediates a fundamental force of nature.}\label{higgsforce}
\end{figure}

\item The Higgs boson mediates a new force of nature, just as the graviton
mediates the gravitational force and the gauge bosons of the standard model
mediate the strong and electroweak forces, as shown in Fig.~\ref{higgsforce}.
Unlike the graviton, which has spin 2, and the gauge bosons, which have spin
1, the Higgs boson has spin 0, since it is a scalar field.

\item The Higgs field is responsible CKM mixing and $CP$ violation.  The
quarks acquire mass via their coupling to the Higgs field,
\begin{displaymath}
{\cal L}_{Yukawa}=\Gamma^{ij}_u\bar Q_L^i\epsilon\phi^*u_R^j+\Gamma^{ij}_d\bar
Q_L^i\phi d_R^j
\end{displaymath}
($Q_L$ is defined in Table~\ref{tab:sm}) where the indicies $i,j=1,2,3$
indicate the generation. The fact that the Yukawa matrices $\Gamma_u,\Gamma_d$
are nondiagonal leads to CKM mixing, and the fact that they are complex yields
$CP$ violation.  An analogous mechanism leads to MNS mixing in the lepton
sector (evidenced via neutrino oscillations), as well as leptonic $CP$
violation (yet to be observed).  We would like to understand the curious
pattern of fermion masses and mixing observed in nature.

\item We don't understand why the cosmological constant is so much less than
the vacuum-expectation value of the Higgs field, $\Lambda \ll v^2$.  One might
imagine that there is some mechanism that forces it to zero, but then we have
to explain why it is observed to be nonzero.  ``Quintessence'' is another
scalar field introduced to provide such an explanation \cite{Carroll:2001xs}.

\item The WMAP measurements provide support for inflation, but we do not know
the dynamics that drive inflation.  Yet another scalar field, the ``inflaton,''
has been proposed for that purpose.

\item As discussed above, gauge-coupling unification relies on the Higgs field (or
fields).  There may be yet more Higgs fields responsible for spontaneously
breaking the grand-unified symmetry.

\end{itemize}

These observations show that the Higgs boson is not only central to the
standard model, it is central to physics beyond the standard model.  The LHC
promises to open up an entirely new chapter in our quest to understand nature
at a deeper level.

\section{The Road Ahead}

We still have a long road ahead of us, but it is worth the wait.  As we
approach our destination, we will encounter a landscape that is familiar in
some ways, exotic in others.  Recall that when the CERN $Sp\bar pS$ first
began operation, there was a lot of confusion: monojets, a 40 GeV top quark,
and so on.  I believe that when we begin the operation of the LHC, the
situation will be both confusing and exhilarating.  It will require the best
efforts of us all to make sense of it.

\section*{Acknowledgments}
I am grateful for conversations and correspondence with K.~Babu,
G.~B\'elanger, A.~Belyaev, J.~Ellis, T.~Liss, K.~Matchev, D.~O'Neil, C.~Quigg,
J.~Richman, M.~Srednicki, and J.~Thaler.  This research was supported in part
by the U.~S.~Department of Energy under contract No.~DE-FG02-91ER40677 and by
the National Science Foundation under Grant No.~PHY99-07949.

%
\end{document}